\documentclass{article}

\usepackage{amsmath,graphicx}

\begin{document}

\title{Finite sampling interval effects in Kramers-Moyal analysis}

\author{Steven J. Lade}

\maketitle

\section*{Abstract}
Large sampling intervals can affect reconstruction of Kramers-Moyal coefficients from data. A new method, which is direct, non-stochastic and exact up to numerical accuracy, can estimate these finite-time effects. For the first time, exact finite-time effects are described analytically for special cases; biologically inspired numerical examples are also worked through numerically. The approach developed here will permit better evaluation of Langevin or Fokker-Planck based models from data with large sampling intervals. It can also be used to predict the sampling intervals for which finite-time effects become significant. 

\section{Introduction}
Kramers-Moyal analysis is a recently developed \cite{Freidrich_PLA_2000} method of characterising, from time series data, an arbitrarily nonlinear stochastic process $X(t)$. The Kramers-Moyal coefficients are defined as
\begin{equation}
\label{eq:KM}
D^{(n)}(x_0,t) = \lim_{\tau \to 0} \frac{1}{\tau n! } \langle [x(t + \tau) - x_0]^n | x(t)=x_0 \rangle.
\end{equation}
The smallest $\tau$ available in an experiment is the sampling interval, but the definition requires an infinitesimally small $\tau$. If the sampling interval is not `small enough', the coefficients measured from experimental data may therefore not match those arising from an analytically constructed continuous model. This article presents, for the first time, a non-stochastic method of exactly predicting these finite sampling interval effects, which may be abbreviated to `finite-time effects' throughout this article.

The first two Kramers-Moyal coefficients appear in Fokker-Planck equations for transition probability density,
\begin{equation}
\label{eq:FP}
\frac{\partial P(x,t|x',t')}{\partial t} =  \hat{L}(x) P(x,t|x',t'),
\end{equation}
where the Fokker-Planck operator is
\begin{equation*}
\hat{L}(x) = -\frac{\partial}{\partial x} D^{(1)}(x) + \frac{\partial^2}{\partial x^2} D^{(2)}(x).
\end{equation*}
These first two coefficients are known as the drift and diffusion coefficients respectively, and are analogous to effective position-dependent force and noise, or `temperature', respectively \cite{Kriso_PLA_2002}. The drift and diffusion coefficients, then, provide an intuitive description of the process. If the process is also Markovian, they provide a complete characterisation; furthermore, all higher coefficients will be zero \cite{Risken}. Unlike many other methods of time series analysis, the Kramers-Moyal approach recovers coefficients with arbitrary nonlinearity, up to the resolution of the bin size chosen in the reconstruction. 

The Kramers-Moyal approach has previously been applied to neuroscience \cite{Freidrich_PLA_2000}, cardiology \cite{Kuusela_PRE_2004}, traffic engineering \cite{Kriso_PLA_2002}, finance \cite{Friedrich_PRL_2000, Ghasemi_PRE_2007} and turbulence \cite{Friedrich_PRL_1997}. These applications frequently assume homogeneity, in the sense that the Kramers-Moyal coefficients are time-invariant, so that the ensemble average in Eq. \eqref{eq:KM} can be replaced by a time average. We will also assume homogeneity throughout this article.

\begin{figure}
\includegraphics[width=9cm]{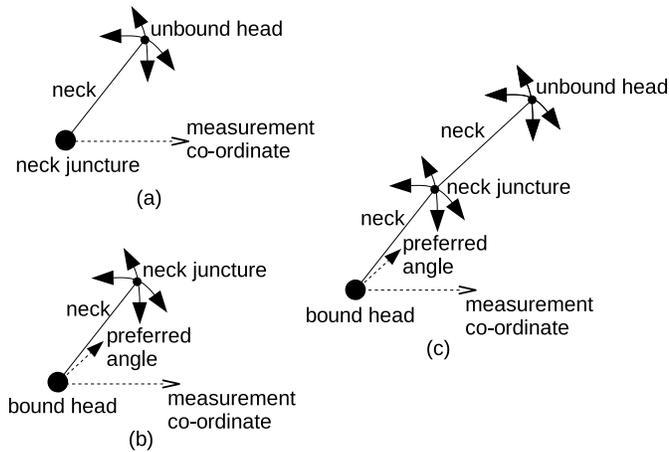}
\caption{Schematic of the biologically-inspired examples used in this article: (a) free diffusion on a sphere, or `tethered diffusion'; (b) biased diffusion on a sphere; and (c) compound diffusion on a sphere.}
\label{fig:models}
\end{figure}

The burgeoning collections of data available with recent advances in quantitative biology also beckons use of such methods, for biological systems are often well-modelled by an overdamped and Brownian environment. For example, one phase in the `walking' of naturally occurring molecular motors such as myosin-V and kinesin is believed to be a tethered diffusion state, where one `head' of the bipedal motor is unbound and searching for the next binding site \cite{Burgess_JCB_2002,Dunn_NSMB_2007,Shiroguchi_S_2007}. We will use this system to illustrate the results we obtain. The unbound head can be modelled as undergoing diffusion on a sphere with center on the neck juncture and with radius equal to the length of the tether. We refer to this case as `free diffusion on a sphere'. However, the neck juncture itself is not fixed, for its tether is flexible about its preferred orientation \cite{Vilfan_BJ_2005}. We model the motion of the juncture as diffusion on a sphere with a potential: `biased diffusion on a sphere'. Combining these models, a more precise treatment of the unbound head would have it undergoing free diffusion on a sphere about a point which is itself undergoing biased diffusion on a sphere: `compound diffusion'. In all cases we assume the experiment can only measure in one linear dimension, inclined at angle $\theta_0$ from the direction of potential minimum (where there is a potential). These are summarised in Fig. \ref{fig:models}.

Ragwitz and Kantz \cite{Ragwitz_PRL_2001} were the first to consider finite sampling interval effects in the estimation of Kramers-Moyal coefficients. They calculated a partial \cite{Friedrich_PRL_2002} approximation to such effects to first order in the sampling interval. An exchange with Friedrich et al. ensued \cite{Friedrich_PRL_2002,Ragwitz_PRL_2002}, with Freidrich et al. providing an infinite series expansion for the finite-time correction. Later, Kleinhans et al. \cite{Kleinhans_PLA_2005} described a numerical method for recovering the continuous-time Kramers-Moyal coefficients from finite-time estimates. From initial continuous-time guesses, their method computed the resulting finite-time estimates by direct numerical simulation of the Langevin equation corresponding to the Fokker-Planck equation \eqref{eq:FP}. They update the guesses based on the known finite-time measurements. Our method avoids the need for stochastic Langevin simulations. 

In Section 2 we review Friedrich et al.'s infinite series formula, then proceed to our main result. Section 3 calculates the finite-time effects for some analytical special cases, while section 4 provides some numerical examples. These examples are chosen for their relevance to possible biophysical uses of Kramers-Moyal analysis. Points of a general nature are made in section 5, and concluding remarks in section 6.

\section{Derivation}
\label{sec:derivation}
We denote the Kramers-Moyal coefficients \eqref{eq:KM} calculated with finite sampling interval $\tau$ by
\begin{equation}
\label{eq:KMtau}
D_\tau^{(n)}(x_0) \equiv \frac{1}{\tau n!} \langle [x(t + \tau) - x_0]^n | x(t)=x_0 \rangle.
\end{equation}
This can also be written as
\begin{equation*}
D_\tau^{(n)}(x_0) \equiv \frac{1}{\tau n!} \int_{-\infty}^\infty (x-x_0)^n P(x,t_0+\tau|x_0,t_0) dx,
\end{equation*}
which since we assume homogeneity is independent of $t_0$. Substituting the formal solution to the Fokker-Planck equation \eqref{eq:FP}
\begin{equation*}
P(x,t+\tau|x_0,t) = e^{\hat{L}(x) \tau} \delta(x-x_0),
\end{equation*}
we obtain
\begin{align}
\label{eq:directL} D_\tau^{(n)}(x_0) &= \frac{1}{\tau n!} \int_{-\infty}^\infty (x-x_0)^n e^{\hat{L}(x) \tau} \delta(x-x0) dx \\
\label{eq:Friedrich} &= e^{\hat{L}^\dagger(x) \tau} (x-x_0)^n|_{x=x_0}.
\end{align}
Here
\begin{equation*}
\hat{L}^\dagger(x) = D^{(1)}(x) \frac{\partial}{\partial x} + D^{(2)} \frac{\partial^2}{\partial x^2}
\end{equation*}
is the adjoint of the operator $\hat{L}$ over the inner product $(g,h) = \int_{-\infty}^\infty g(x) h(x) dx$. Friedrich et al. \cite{Friedrich_PRL_2002} expand the symbolic exponential, and show that the first-order terms lead to the first-order correction of Ragwitz and Kantz \cite{Ragwitz_PRL_2001}.

Equation \eqref{eq:Friedrich} has an alternative interpretation, by a route Zwanzig \cite{Zwanzig_2001} calls the `Heisenberg approach' in analogy to the complementary formulations of quantum mechanics. This interpretation is that the solution of the partial differential equation
\begin{equation}
\label{eq:adjointFP} \frac{\partial W(x,t)}{\partial t} = L^\dagger(x) W(x,t)
\end{equation}
at $(x_0,\tau)$, with initial condition $W(x,0) = (x-x_0)^n/n! \tau$, gives $D_\tau^{(n)}(x_0)$. 

Therefore, knowing the continuous-time $D^{(1)}(x)$ and $D^{(2)}(x)$ we can predict the effect of finite sampling interval without direct simulation of the Langevin equations. This is the main result of this article. We will refer to Eq. \eqref{eq:adjointFP} as the adjoint Fokker-Planck equation; note it is different to, and should not be confused with, the Kolmolgorov backward equation \cite{Risken}.

One other obvious non-stochastic approach is to compute Eq. \eqref{eq:directL}, but this involves a Dirac-delta initial condition. For some special cases, this initial condition could be treated analytically. Indeed, this is the method by which Ragwitz and Kantz \cite{Ragwitz_PRL_2001} obtained their approximate corrections. In general, however, where numerical methods are required, the Dirac-delta initial condition would present significant difficulties.

\section{Analytic special cases}
In some cases we can solve the adjoint Fokker-Planck equation \eqref{eq:adjointFP} analytically. One of the simplest scenarios is a linear drift $D^{(1)}(x) = -kx$, corresponding to a quadratic potential well, and a constant diffusion $D^{(2)}(x) = D$. This is the Ornstein-Uhlenbeck process in one dimension \cite{Risken}. With initial condition $W(x,0) = x - x_0$, the solution is $W(x,\tau) = xe^{-k\tau}-x_0$, so the finite-time drift coefficient measured at finite sampling interval $\tau$ is
\begin{equation}
\label{eq:OUdrift}
D_\tau^{(1)}(x) = -\frac{x}{\tau}(1-e^{-k\tau}).
\end{equation}
With initial condition $W(x,0) = (x - x_0)^2$, the solution is $W(x,\tau) = (xe^{-k\tau}-x_0)^2 + \frac{D}{k}(1-e^{2k\tau})$. Therefore the finite-time diffusion coefficient is
\begin{equation}
\label{eq:OUdiffusion}
D_\tau^{(2)}(x) = \frac{1}{2\tau}\left[x^2(1-e^{-k\tau})^2 + \frac{D}{k}(1-e^{-2k\tau})\right].
\end{equation}

As $\tau$ increases, the finite-time drift coefficient for the 1-D Ornstein-Uhlenbeck process remains linear but decreases in gradient, while a quadratic component appears in the diffusion. This causes the diffusion estimate to be larger at the edges of a potential well than at the center, an effect first predicted by Ragwitz and Kantz \cite{Ragwitz_PRL_2001}.

Free diffusion on the surface of a sphere of radius $r$, as sketched in Fig. \ref{fig:models}(a), can be represented in polar co-ordinates $(\theta,\phi)$ by the Langevin equations \cite{Raible_AOC_2004}
\begin{align*}
d\theta &= D \cot \theta dt + \sqrt{2 D} d w_\theta \\
d\phi &= \frac{\sqrt{2 D}}{\sin\theta} dw_\phi,
\end{align*}
written in the It\=o convention. Projected onto a rectangular co-ordinate $x = r\cos\theta$, these reduce to the Langevin equation \cite{Lade_geom_inprep}
\begin{equation}
\label{eq:freediff}
dx = -2Dx dt - \sqrt{2D(r^2 - x^2)} dw,
\end{equation}
corresponding to drift and diffusion coefficients $D^{(1)}(x) = -2Dx$ and $D^{(2)}(x) = D(r^2 - x^2)$. With initial condition $W(x,0) = x - x_0$, the solution to the adjoint Fokker-Planck equation \eqref{eq:adjointFP} is $W(x,\tau) = xe^{-2D\tau}-x_0$, so the finite-time drift coefficient is
\begin{equation}
\label{eq:spheredrift}
D_\tau^{(1)}(x) = -\frac{x}{\tau}(1-e^{-2D\tau}).
\end{equation}
With initial condition $W(x,0) = (x - x_0)^2$, the solution is $W(x,\tau) = x^2 e^{-6D\tau} - 2x_0 x e^{-2D\tau} + x_0^2 - r^2 \left(e^{-6D\tau} - 1 \right)/3$. Therefore the finite-time diffusion coefficient is
\begin{equation}
\label{eq:spherediffusion}
D_\tau^{(2)}(x) = \frac{1}{2\tau}\left[x^2 \left(1+e^{-6D\tau}-2e^{-2D\tau} \right) + \left(1-e^{-6D\tau}\right)\right].
\end{equation}

For free diffusion on a sphere, the gradient of the drift coefficient \eqref{eq:spheredrift} again simply decreases with $\tau$. Notice that the diffusion \eqref{eq:spherediffusion}, however, undergoes a much more qualitatively significant change: it changes from convex to concave as $\tau$ increases. This change occurs at $\tau = \ln(1/2+\sqrt{5}/2)/2D \approx 1/4D$.

As a general rule of thumb, we see that if the sampling interval is of order $1/k$ (for Ornstein-Uhlenbeck) or $1/D$ (free diffusion on a sphere), or larger, we should be concerned about finite sampling interval effects. Also, a concave diffusion is an indicator there may be finite-time effects. In both examples the original $D^{(1)}$ and $D^{(2)}$ are recovered in the limit $\tau \to 0$.

\section{Numerics}
\label{sec:numerics}
The finite-time $D^{(n)}_\tau(x_0)$ may also be computed numerically. For each value $D^{(n)}_\tau(x_0)$ that is required, the adjoint Fokker-Planck equation \eqref {eq:adjointFP} with initial condition $W(x,0) = (x-x_0)^n/n! \tau$ must be solved up to time $t = \tau$.

\begin{figure}
\includegraphics[width=9cm]{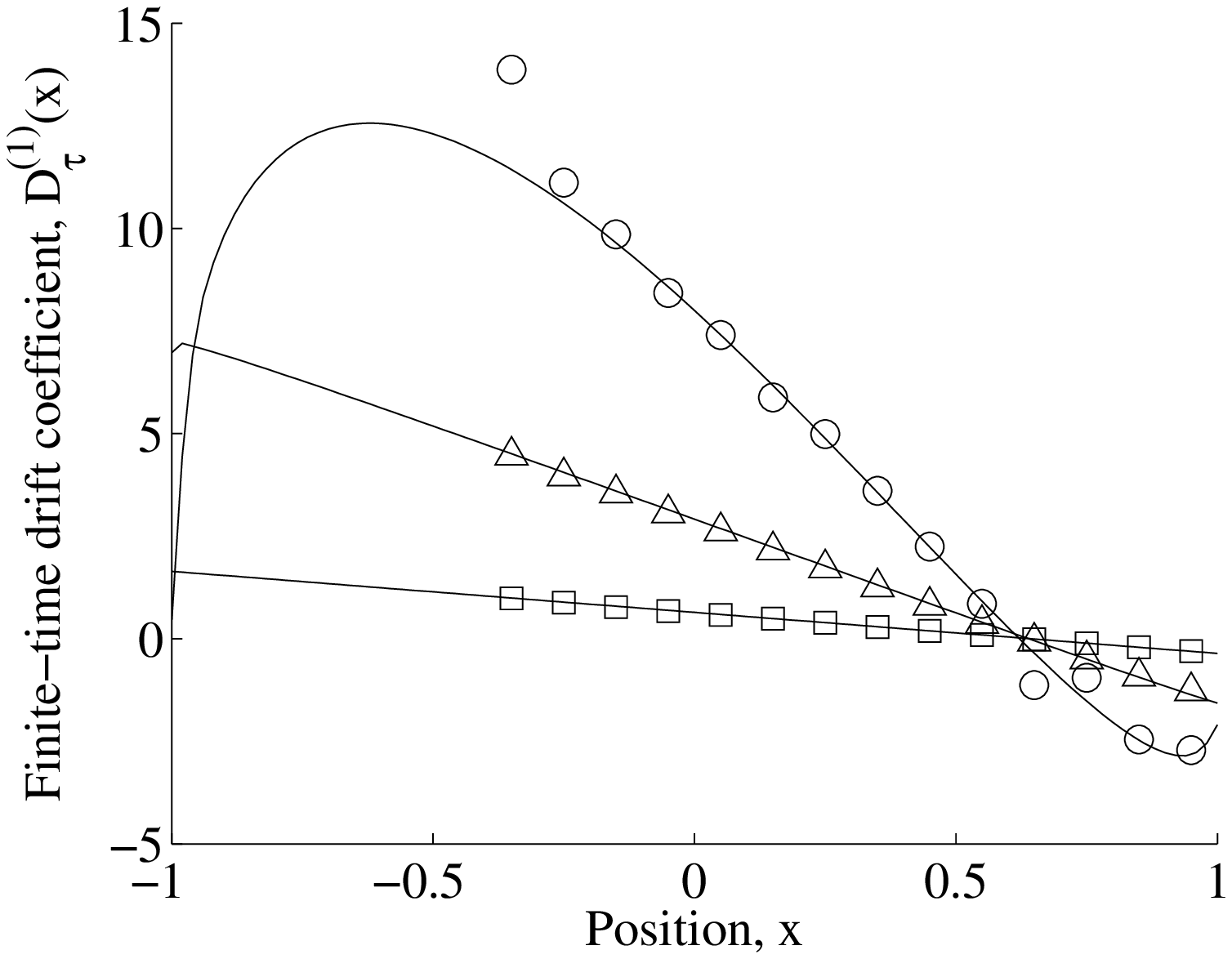} \includegraphics[width=9cm]{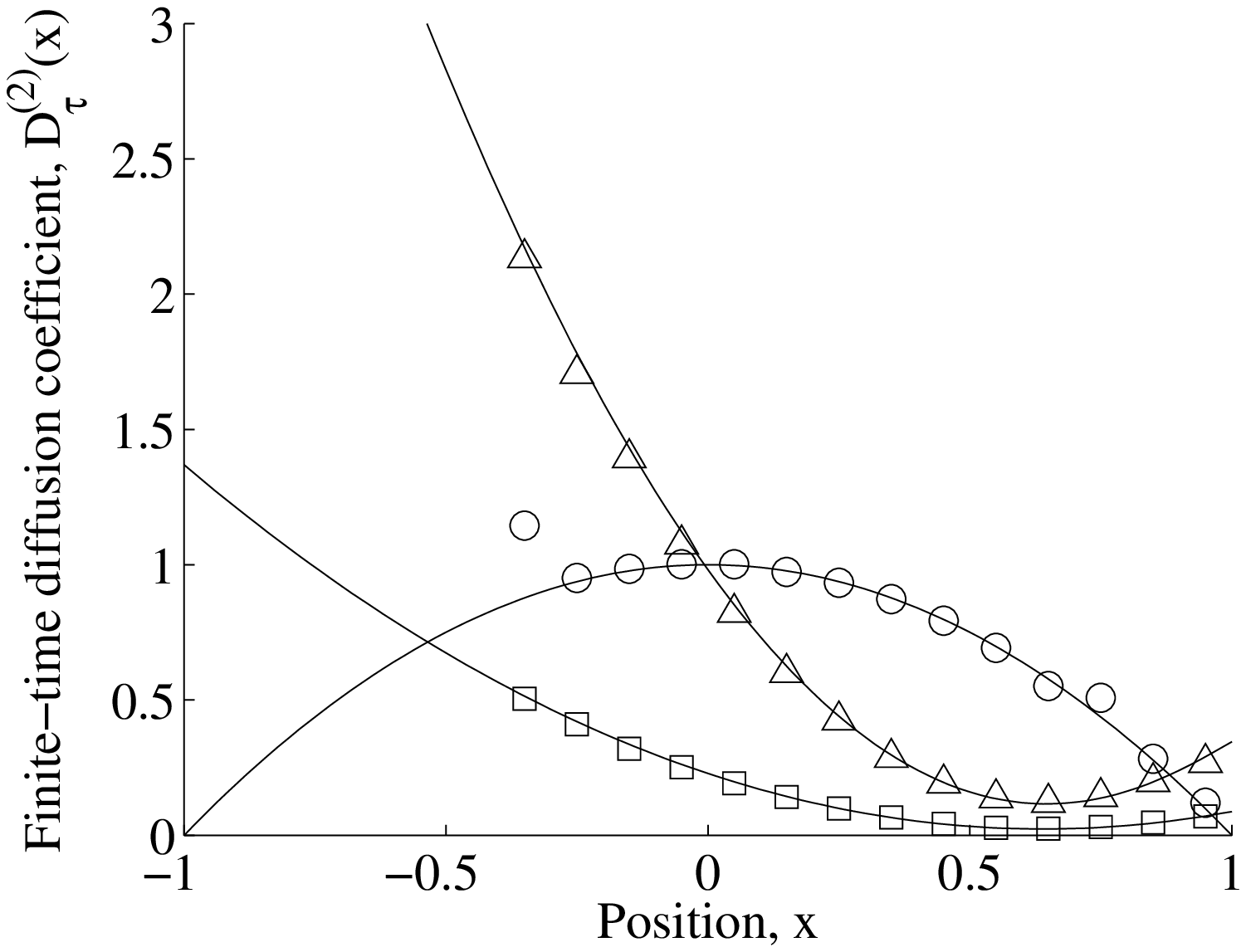}
\caption{Finite-time (top) drift and (bottom) diffusion coefficients for biased diffusion on a sphere, predicted semi-analytically from the method of section \ref{sec:derivation} (solid lines) and by direct numerical simulation (markers). The parameters were $k = 1$, $\theta_0 = \pi/4$, $D = 1$ and sampling interval $\tau \to 0$ (circles), $\tau = 1$ (triangles) and $\tau = 5$ (squares). The direct numerical results were not plotted where there were insufficient data.}
\label{fig:biaseddiff}
\end{figure}

Biased diffusion on a sphere [Fig. \ref{fig:models}(b)] can be modelled by the Langevin equations, in polar co-ordinates $(\theta, \phi)$, by \cite{Raible_AOC_2004}
\begin{align}
\label{eq:biaseddiff}
\begin{split}
d\theta &= (-\partial_\theta U + D \cot \theta) dt + \sqrt{2 D} d w_\theta \\
d\phi &= -\frac{\partial_\phi U}{\sin^2 \theta} dt + \frac{\sqrt{2 D}}{\sin\theta} dw_\phi.
\end{split}
\end{align}
We use a harmonic potential $U = k\theta^2/2$; for $k = 0$ we recover the free diffusion on a sphere of the previous section. We projected the system onto one-dimensional continuous-time Kramers-Moyal coefficients by the method of Lade and Kivshar \cite{Lade_geom_inprep}. This co-ordinate was the rectangular co-ordinate at angle $\theta_0$ from the preferred angle of orientation ($\theta = 0$), that is, $x = \cos\theta_0 \cos\theta + \sin\theta_0 \sin\theta \cos\phi$. We then estimated the finite-time $D^{(1)}_\tau(x)$ and $D^{(2)}_\tau(x)$ for a variety of $\tau$ by numerically solving the adjoint Fokker-Planck equation with a simple forward-time centred-space scheme and extrapolated boundary conditions. For comparison, we also computed the drift and diffusion coefficients, as per Eq. \eqref{eq:KMtau}, from direct simulation of the Langevin equations \eqref{eq:biaseddiff}. The results are shown in Fig. \ref{fig:biaseddiff}.

We observe the same decrease in gradient of the drift as in the Ornstein-Uhlenbeck process. Since the drift is roughly displacement divided by sampling interval $\tau$, and since the displacement of the projection $x$ between measurements is limited (by the radius of the sphere), it makes sense that the estimated drift decreases with increasing $\tau$. Unlike the Ornstein-Uhlenbeck process, there is some curvature due to geometrical effects. The zero crossing point reflects the position of the potential minimum at $x = \cos\theta_0$. With the diffusion, we observe a loss of features and convergence to an upright quadratic shape as in the Ornstein-Ulhenbeck process. As observed by Ragwitz and Kantz \cite{Ragwitz_PRL_2001}, this shape is because at large $\tau$, nonzero drift can lead to overestimation of diffusion, and the drift is larger at large position $x$. In both drift and diffusion curves, there are errors in the direct numerical estimates due to the singularities in the Langevin equation \eqref{eq:biaseddiff} at $\theta = 0$ and $\pi$.

\begin{figure}
\includegraphics[width=9cm]{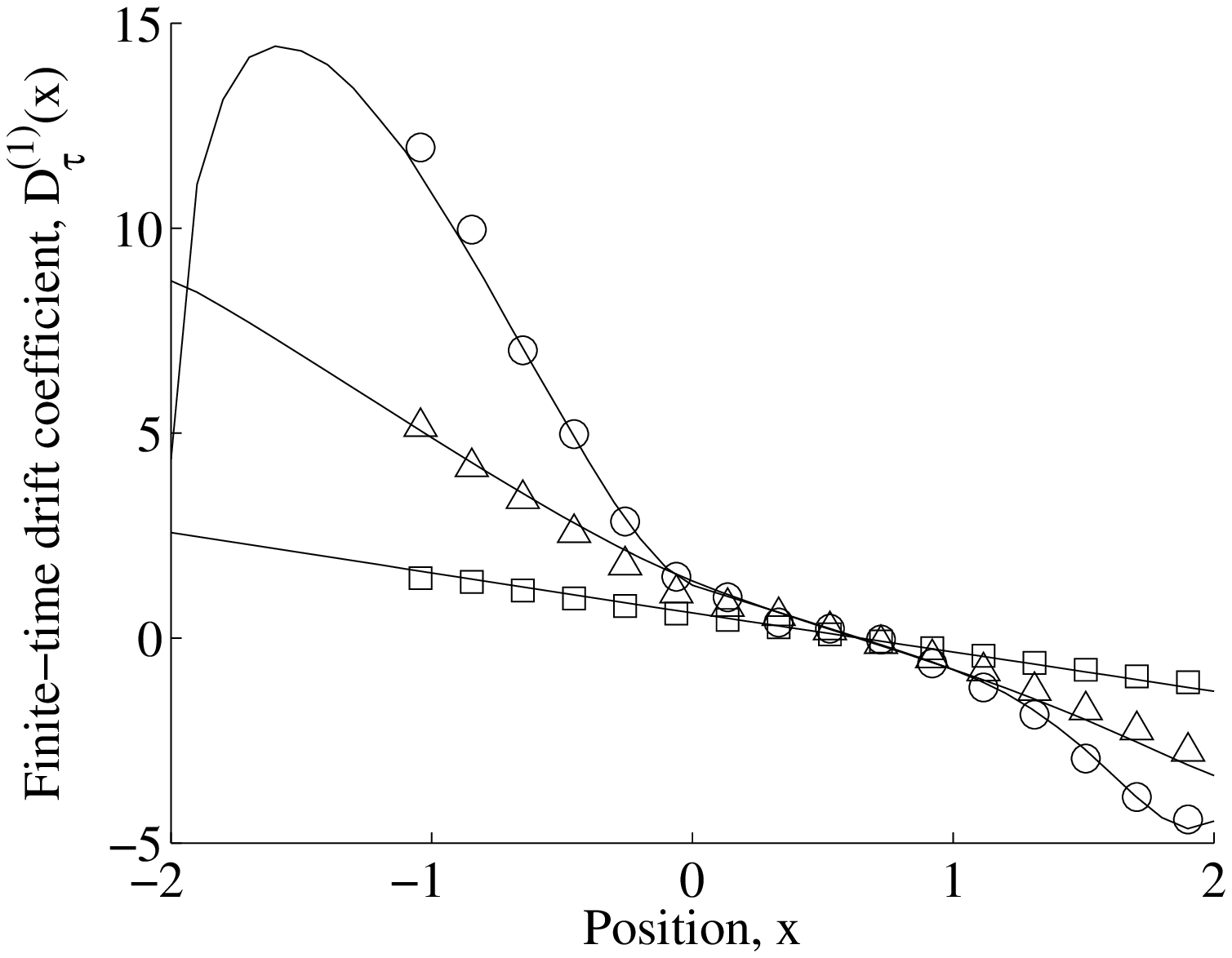} \includegraphics[width=9cm]{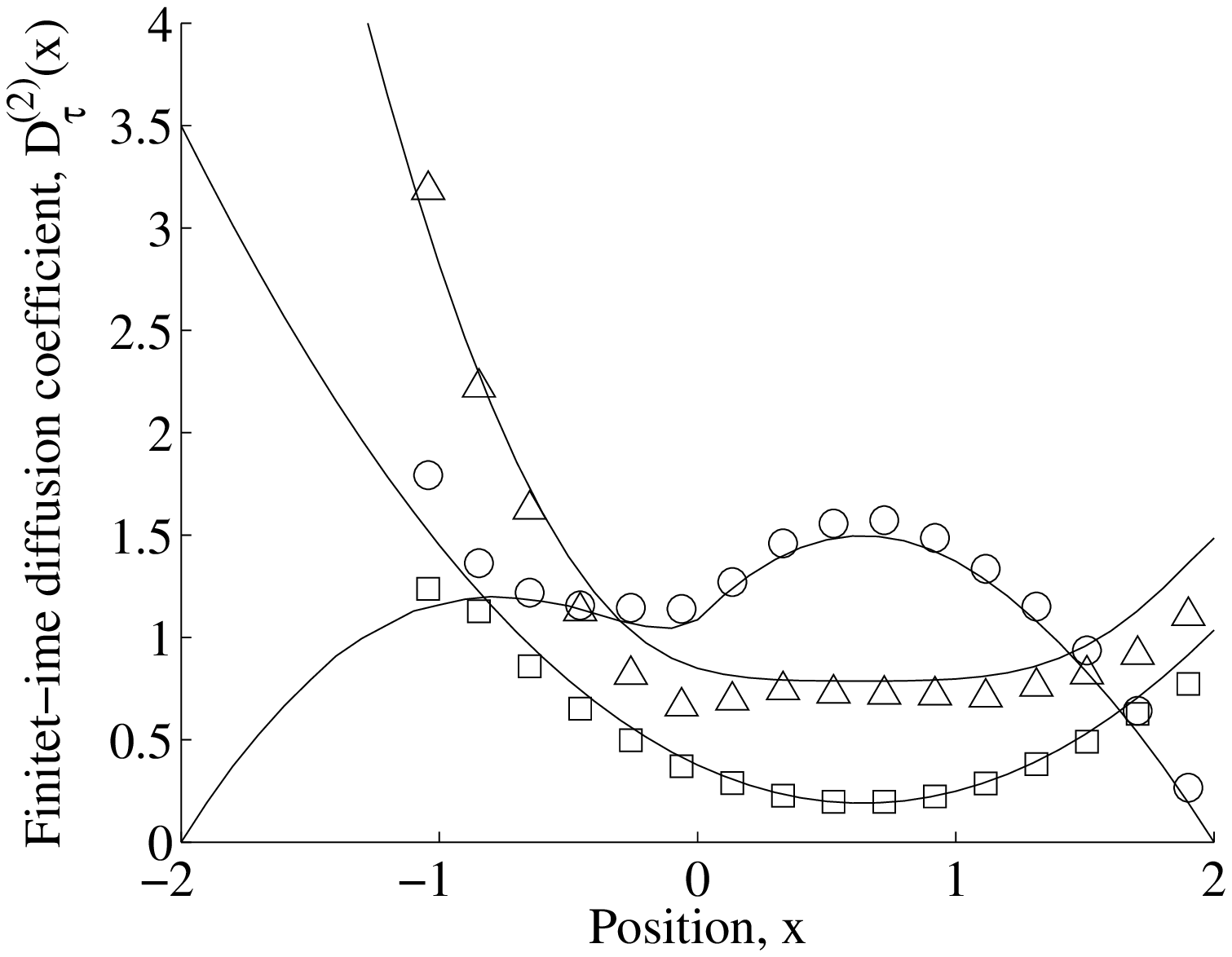}
\caption{Finite-time (top) drift and (bottom) diffusion coefficients for compound diffusion on a sphere, predicted semi-analytically from the method of section \ref{sec:derivation} (solid lines) and by direct numerical simulation (markers). The parameters were $k = 1$, $\theta_0 = \pi/4$ (for the biased diffusion), $D = 1$ (for both parts of the compound) and sampling interval $\tau \to 0$ (circles), $\tau = 1$ (triangles) and $\tau = 5$ (squares). The direct numerical results were not plotted where there were insufficient data.}
\label{fig:compound}
\end{figure}

As a second example, we simulate compound diffusion on a sphere [Fig. \ref{fig:models}(c)], where a point undergoes free diffusion on a sphere \eqref{eq:freediff} with respect to a point that is itself undergoing biased diffusion on a sphere \eqref{eq:biaseddiff}. Results are in Fig. \ref{fig:compound}.

Here we observe some interesting geometric effects \cite{Lade_geom_inprep} in the continuous-time ($\tau \to 0$) drift and diffusion coefficients but which become `washed out' once again to linear drift and upright quadratic diffusion for large $\tau$.

\section{Discussion}
The characteristic quadratic shape we have obtained for the diffusion coefficient at large sampling intervals is clearly observable in Farapour et al. \cite{Farahpour_PA_2007} and Ghasemi et al. \cite{Ghasemi_PRE_2007}, and was used by them in fitting the results of their analyses. If we assumed their process was actually Ornstein-Uhlenbeck, using their sampling intervals and the analytic solutions (\ref{eq:OUdrift}-\ref{eq:OUdiffusion}) we could solve for $k$ and $D$ in each case, and found a good match with their numerical drift and diffusion curves. Note that, given $\tau$, this problem is overdetermined: the functional forms leave three simultaneous equations from which to determine $k$ and $D$. Thus the processes investigated by those authors (specfically, the logarithmic increments) are well modelled in continuous time by the above simple Ornstein-Uhlenbeck process.

The drift and diffusion coefficients observed in some cascade analyses of time series \cite{Kimiagar_JSM_2008,Jafari_PRL_2003,Friedrich_PRL_2000} also display patterns similar to the above, with the scale size standing in for sampling interval. For Jafari et al. \cite{Jafari_PRL_2003} the trends match almost exactly with $k = 0.0055$, $D = 2.9 \times 10^{-4}$ and $\tau = \Delta x / 8$ with length scale $\Delta x$ given as in the legends to their figure 2. For Kimiagar et al. \cite{Kimiagar_JSM_2008} the directions of the trends do not match as precisely, while Friedrich et al. \cite{Friedrich_PRL_2000} only graph for one $\tau$. In these two cases the finite-time drift and diffusion coefficients seem to match those expected for the Ornstein-Uhlenbeck model above, but there is no explicable mapping from length scale $\Delta x$ onto $\tau$.

Several of these authors \cite{Jafari_PRL_2003,Kimiagar_JSM_2008,Ghasemi_PRE_2007} then perform Langevin simulations with their numerical drift and diffusion coefficients, which we have shown to be suffering from finite-time effects. But in all cases they observe excellent agreement with the reference time series. We claim this is because the {\it finite-time} drift and diffusion coefficients in general provide for better simulation of the time series if {\it the same time step as the original sampling interval} is used, a subtlety these authors overlook. For example, in the simple Euler-Maryuama \cite{KloedenPlaten_1995} algorithm, $D_\tau^{(1)}$ by construction provides an unbiased estimate for $x(t+\tau)$ from $x(t)$.

We have assumed that the evolution of the measurand can be fully described by the Fokker-Planck equation \eqref{eq:FP} and the drift and diffusion coefficients. This is valid if the process is Markovian. Projections of a larger-dimensional system, such as those used in section 4, are in general not Markov, even when the original system is \cite{Zwanzig_2001}. However, it has been shown that the projections of section \ref{sec:numerics} are approximately Markovian \cite{Lade_geom_inprep}, and the excellent agreement with direct numerical simulations in that section confirms that the method worked successfully. Also, the method could easily be extended to other evolution operators $\hat{L}$ than the Fokker-Planck operator, in the event that this operator is not appropriate.

In principle, the numerical method outlined above could be inverted iteratively, preferably with more sophisticated methods of numerical solution, to find continuous drift and diffusion coefficients from experimentally measured finite-time coefficients. The only published method \cite{Kleinhans_PLA_2005} to solve this `inverse problem' is also numerical, but relies instead on direct simulation of the estimated system's Langevin equations of motion, which being stochastic must be repeated over many trajectories to acheive good statistics.

Even in the absence of such modifications, the results above provide worthwhile contributions to time series analysis by Kramers-Moyal coefficients. For the first time without direct integration of the Langevin equation of motion, we have provided an exact method with which to predict the effects of finite sampling interval on the estimation of Kramers-Moyal coefficients. Our method contains no stochastic terms, being based on Fokker-Planck equations, and requires no ensemble (or time) averaging. 

Given a known model for a system, our approach can be used to predict the effect of finite experimental sampling intervals on the estimation of Kramers-Moyal coefficients. Experimental observations of the Kramers-Moyal coefficients could then be used to support or reject the model.

Given a time series, a quick test to see whether finite-time effects may occur could be to check whether the sampling interval approaches any characteristic periods of the system. This would occur if the differences in the time series over a single time step are `large' compared to the full extent of the time series' fluctuations.

\section{Conclusions}
An alternative interpretation of an existing formula \eqref{eq:Friedrich} was proposed, which permits exact prediction of finite sampling interval effects on the estimation of Kramers-Moyal coefficients. Special analytical cases were presented, which showed general features of finite sampling interval effects. The approach was also implemented numerically, for examples of particular relevance to biophysics. Previously published Kramers-Moyal analyses showed features possibly explicable as finite-time effects, in particular, concave diffusion curves. The work can permit better evaluation of Langevin or Fokker-Planck based models with data that has large sampling intervals, or to predict the sampling intervals for which finite-time effects should become significant.

\bibliographystyle{elsarticle-num}
\bibliography{myosin}

\begin{thebibliography}{10}
\expandafter\ifx\csname url\endcsname\relax
  \def\url#1{\texttt{#1}}\fi
\expandafter\ifx\csname urlprefix\endcsname\relax\def\urlprefix{URL }\fi
\expandafter\ifx\csname href\endcsname\relax
  \def\href#1#2{#2} \def\path#1{#1}\fi

\bibitem{Freidrich_PLA_2000}
R.~Friedrich, S.~Siegert, J.~Peinke, S.~L\"uck, M.~Siefert, M.~Lindemann,
  J.~Raethjen, G.~Deuschl, G.~Pfister, Extracting model equations from
  experimental data, Phys. Lett. A. 271 (2000) 217--222.

\bibitem{Kriso_PLA_2002}
S.~Kriso, J.~Peinke, R.~Friedrich, P.~Wagner, Reconstruction of dynamical
  equations for traffic flow, Phys. Lett. A 299~(2-3) (2002) 287--291.

\bibitem{Risken}
H.~Risken, The Fokker Planck Equation: Methods of Solution and Applications,
  Springer-Verlag, Berlin, 1984.

\bibitem{Kuusela_PRE_2004}
T.~Kuusela, Stochastic heart-rate model can reveal pathologic cardiac dynamics,
  Phys. Rev. E 69~(3) (2004) 031916.

\bibitem{Friedrich_PRL_2000}
R.~Friedrich, J.~Peinke, C.~Renner, How to quantify deterministic and random
  influences on the statistics of the foreign exchange market, Phys. Rev. Lett.
  84~(22) (2000) 5224--5227.

\bibitem{Ghasemi_PRE_2007}
F.~Ghasemi, M.~Sahimi, J.~Peinke, R.~Friedrich, G.~R. Jafari, M.~R.~R. Tabar,
  Markov analysis and {Kramers-Moyal} expansion of nonstationary stochastic
  processes with application to the fluctuations in the oil price, Phys. Rev. E
  75~(6) (2007) 060102.

\bibitem{Friedrich_PRL_1997}
R.~Friedrich, J.~Peinke, Description of a turbulent cascade by a
  {Fokker-Planck} equation, Phys. Rev. Lett. 78~(5) (1997) 863--866.

\bibitem{Burgess_JCB_2002}
S.~Burgess, M.~Walker, F.~Wang, J.~R. Sellers, H.~D. White, P.~J. Knight,
  J.~Trinick, The prepower stroke conformation of myosin {V}, J. Cell Biol.
  159~(6) (2002) 983--991.

\bibitem{Dunn_NSMB_2007}
A.~R. Dunn, J.~A. Spudich, Dynamics of the unbound head during myosin {V}
  processive translocation, Nature Struct. Mol. Biol. 14 (2007) 246--248.

\bibitem{Shiroguchi_S_2007}
K.~Shiroguchi, J.~Kinosita, Kazuhiko, Myosin {V} walks by lever action and
  {B}rownian motion, Science 316~(5828) (2007) 1208--1212.

\bibitem{Vilfan_BJ_2005}
A.~Vilfan, Elastic lever-arm model for myosin {V}, Biophys. J. 88~(6) (2005)
  3792 -- 3805.

\bibitem{Ragwitz_PRL_2001}
M.~Ragwitz, H.~Kantz, Indispensable finite time corrections for {Fokker-Planck}
  equations from time series data, Phys. Rev. Lett. 87~(25) (2001) 254501.

\bibitem{Friedrich_PRL_2002}
R.~Friedrich, C.~Renner, M.~Siefert, J.~Peinke, Comment on `{I}ndispensable
  finite time corrections for {F}okker-{P}lanck equations from time series
  data', Phys. Rev. Lett. 89~(14) (2002) 149401.

\bibitem{Ragwitz_PRL_2002}
M.~Ragwitz, H.~Kantz, Ragwitz and {K}antz reply, Phys. Rev. Lett. 89~(14)
  (2002) 149402.

\bibitem{Kleinhans_PLA_2005}
D.~Kleinhans, R.~Friedrich, A.~Nawroth, J.~Peinke, An iterative procedure for
  the estimation of drift and diffusion coefficients of {Langevin} processes,
  Phys. Lett. A 346~(1-3) (2005) 42 -- 46.

\bibitem{Zwanzig_2001}
R.~Zwanzig, Nonequilibrium Statistical Mechanics, Oxford University Press, New
  York, 2001.

\bibitem{Raible_AOC_2004}
M.~Raible, A.~Engel, Langevin equation for the rotation of a magnetic particle,
  Appl. Organometal. Chem. 18 (2004) 536--541.

\bibitem{Lade_geom_inprep}
S.~J. Lade, Y.~S. Kivshar, Geometric and projection effects in
  {K}ramers-{M}oyal analysis, submitted.

\bibitem{Farahpour_PA_2007}
F.~Farahpour, Z.~Eskandari, A.~Bahraminasab, G.~Jafari, F.~Ghasemi, M.~Sahimi,
  M.~R.~R. Tabar, A {L}angevin equation for the rates of currency exchange
  based on the {M}arkov analysis, Physica A 385~(2) (2007) 601--608.

\bibitem{Kimiagar_JSM_2008}
S.~Kimiagar, G.~R. Jafari, M.~R.~R. Tabar, Markov analysis and
  {K}ramers-{M}oyal expansion of the ballistic deposition and restricted
  solid-on-solid models, J. Stat. Mech. (2008) P02010.

\bibitem{Jafari_PRL_2003}
G.~R. Jafari, S.~M. Fazeli, F.~Ghasemi, S.~M.~V. Allaei, M.~R.~R. Tabar, A.~I.
  zad, G.~Kavei, Stochastic analysis and regeneration of rough surfaces, Phys.
  Rev. Lett. 91~(22) (2003) 226101.

\bibitem{KloedenPlaten_1995}
P.~E. Kloeden, E.~Platen, Numerical solution of stochastic differential
  equations, 2nd Edition, Springer, 1995.

\end{thebibliography}

\end{document}